\documentclass{agujournal}

\journalname{<JGR-Space Physics>}

\begin{document}

%
%


\title{Radial transport and plasma heating in Jupiter's magnetodisc}

%
%




\authors{C. S. Ng\affil{1},  P. A. Delamere\affil{1}, V. Kaminker\affil{1}, P. A. Damiano\affil{1} }


\affiliation{1}{Geophysical Institute, University of Alaska Fairbanks, Fairbanks, Alaska, USA}




\correspondingauthor{Chung-Sang Ng}{cng2@alaska.edu}




\begin{keypoints}
\item Jupiter's observed radial ion temperature profile is found to be consistent with heating by turbulent magnetic
  field fluctuations. 
\item An advective outflow approach for calculating turbulent heating is reasonable beyond 10 ${\rm
    R_J}$ where transport becomes rapid and dominated by large-scale motion. 
\item The turbulent heating rate density has been recalculated, resulting in much higher values 
for the region between 10 and 20 ${\rm R_J}$ than previously reported. This increase is critical in the calculation
of temperature that is consistent with observational data. 
\end{keypoints}

%
%


\begin{abstract}

The ion temperature of the magnetosphere of Jupiter derived from
Galileo PLS data was observed to increase by about an order of
magnitude from 10 to 40 Jupiter radii. This suggests the presence of
heating sources that counteract the adiabatic cooling effect of
expanding plasma. There have been different attempts of explaining
this phenomena, including a magnetohydrodynamic (MHD) turbulent
heating model which is based on flux tube diffusion [Saur,
Astrophys. J. Lett., 602, L137, 2004]. We explore an alternate turbulent heating model 
based on advection, similar to models commonly used in solar wind
heating. Based on spectral analysis of Galileo magnetometer (MAG)
data, we find that observed MHD turbulence could potentially provide
the required heating to explain some of the increase in plasma
temperature. 
This indicates that advection is a more appropriate way to describe radial
transport of plasma in the jovian magnetosphere beyond 10 Jupiter radii.

\end{abstract}

%
%

%


%
%
%
%

\section{Introduction}


Jupiter's immense magnetosphere is due to the combination of a strong
planetary magnetic field, rapid rotation and an internal plasma
source \citep[]{Kivelson14, Delamere15_ssr, Achilleos15}.  Io's supplies neutral gas to the inner magnetosphere at a
rate of roughly 1 ton/second and most of this gas is ionized forming
the Io plasma torus \citep[]{Schneider07}.  However, the accumulation of plasma mass cannot
be sustained due to outward radial plasma transport in the form of a
centrifugally-driven flux tube interchange instability in a dipole
magnetic field configuration
\citep[]{Ma16}.   As the magnetic field becomes stretched into a
magnetodisc configuration, it has been suggested that radial transport
is governed by magnetodisc reconnection in the middle and outer magnetosphere \citep[]{Delamere15}. Empirical evidence shows that the plasma must be heated
nonadiabatically during the transport process \citep[]{Bagenal11}).
For example, in Fig. 12 of \citep[]{Bagenal11}), the quantity $PV^\gamma$ is plotted
for both the thermal and hot particle populations
over the range of 6 to 40 radii of Jupiter ${\rm R_J}$ and further, showing that it is far from 
conserved at Jupiter. 
In this paper we address nonadiabatic heating with a turbulent heating
model based on advective outflow commonly used for solar wind heating  \citep[see
e.g.][and references therein]{Ng10a,Ng10b}.

\citet{Southwood87} showed that if the mass content per
unit magnetic flux, $\eta = \int \left(\rho/B\right) ds$
decreases with radial distance (i.e. $\partial \eta/\partial L < 0$, 
where $L$ is the radial distance in the unit of  ${\rm R_J}$) then the
plasma torus is centrifugally unstable. 
Note that $\rho$ is the mass density and $B$ is the magnetic field strength,
with the integral over the path length $ds$ of the magnetic field line along a flux tube.
Quantities defined similarly are the flux tube volume per unit of magnetic flux $V = \int ds/B$,
and the flux tube integrated number density $N = \int n ds$ with $n$ being the number density of particles.
Also, if the flux tube
  entropy, $S = \int \left(p^{1/\gamma}/B\right) ds$, decreases with radial distance, then the torus is unstable.   However, the
  flux tube entropy increases
  with radial distance due to an ever increasing flux tube volume and
  because of increasing
  plasma pressure in the middle and outer magnetosphere
  \citep[]{Paranicas91, Mauk04}.  In general, if $\partial
  S/\partial L > 0$ then the magnetodisc should be stable to
  centrifugal interchange motion.  Understanding the competing
  effects of negative flux tube mass content gradients and  positive
  flux tube entropy gradients is key to understanding the nature of
  radial transport physics. 

In the inner magnetosphere (e.g., inside of 9 ${\rm R_J}$), the torus
exhibits a strong negative flux tube content gradient
\citep[]{Bagenal94}.  In Fig. 9 of \citep{Bagenal16}, radial profiles of
flux tube content ($NL^2$, $\eta$) are plotted from a recent re-analysis of
Galileo plasma science instrument (PLS) data.
It is shown that beyond 10 ${\rm R_J}$, the gradient of these quantities 
is significantly reduced, which naturally precludes significant diffusion.  It is
also in this region that the radial transport rates increase
significantly.  In Fig. 10 of \citep{Bagenal11}, the empirically-derived
radial transport rates and integrated transport times are plotted.  
It is shown that the integrated transport time
from 6 to 10 ${\rm R_J}$ is 10 to 40 days, while from 10 ${\rm R_J}$  to the outer
magnetosphere is on the order of one day with radial outflow speeds
$\sim$10s to 100s km/s.  \citet{Delamere10} noted that the radial outflow speed
can be comparable to the local Alfv\'{e}n speed (based on the
unperturbed dipole) in the middle to outer magnetosphere, suggesting
that the concept of a ``planetary wind'' analogous to the situation in
the solar wind is possible \citep[]{Kennel77}.  Alternatively, truncation of
Alfv\'{e}nic communication to the planet can also be achieved via
pressure anisotropy \citep[]{Kivelson05,Vogt14}, allowing for a free
(e.g., ballooning-type)
outflow.  Thus, we suggest that radial transport beyond 10 ${\rm R_J}$
is governed by magnetodisc reconnection in an impulsive manner
(i.e,. bursty bulk flows) and
involving large-scale radial motions of mixed entropy flux tubes
\citep{Delamere11}.  Therefore, we suggest that while slow diffusive
transport is valid in the inner magnetosphere ($<$ 10 ${\rm R_J}$),
transport in the middle magnetosphere might be better modeled with large-scale
convective inflow/outflow channels. 

\citet{Saur04} considered the possibility of turbulent heating of
Jupiter's magnetodisc (10 to 40 ${\rm R_J}$) using a diffusive
transport model.  The diffusion equation for a dipole geometry is of the form
\begin{equation}\label{equ:diffusion}
    \frac{\partial Y}{\partial t}= L^2 \frac{\partial}{\partial L}\left( \frac{D_{LL}}{L^2} \frac{\partial Y}{\partial L}\right)  \sim 0 \; ,
 \end{equation}  
for quasi steady state, where  $L$ is the radial coordinate ($L = R/R_J$ where ${\rm R_J}$ is the
planetary radius) and $Y$ is any conserved quantity during flux tube interchange
motion.  The conserved quantity for mass is $Y = NL^2$, or the total
number of ions
per unit of magnetic flux. For energy, the flux tube content for a
centrifugally confined plasma is $NL^2 3k_BT/2$, with $k_B$ being the Boltzmann constant.
Following \citet{Richardson83},
the diffusion equation for the flux tube energy is 
\begin{equation}\label{equ:e-diffusion}
     L^2 \frac{\partial}{\partial L}\left( \frac{D_{LL}}{L^2} \frac{3\partial nk_BTL^{3+\gamma}}{2\partial L}\right)  = -qZR^2_JL^{3+\gamma}\; ,
 \end{equation}  
where $\gamma =2$ is used for a equatorially confined plasma sheet with a hight of $Z \sim 2R_J$.
The $q$ factor in the right hand side of Eq.~(\ref{equ:e-diffusion}) is the per volume heating rate, 
presumably from turbulent heating in this model.
Assuming $n = n_0L^{-\beta}$, with $\beta \sim 6.6$ from observation, and 
$D_{LL}=D_0L^{\beta}$ at the same time, Eq.~(\ref{equ:e-diffusion}) can be integrated to obtain an 
equation for the temperature,
\begin{equation}\label{equ:temp-integrated}
     T = \left(T_0 - \frac{F_E}{F_n}\right)\left(\frac{L}{L_0}\right)^{\beta - 3 -\gamma} 
     +  \frac{F_E}{F_n}\left(\frac{L}{L_0}\right)^{-\gamma} + \frac{4\pi \left(3-\beta\right)Z_0 R^2_J}
     {3 k_B F_n  L^{-\beta + 3 +\gamma}} \int^L_{L_0 }dL' L'^{2-\beta}
     \int^{L'}_{L_0}dL'' q L''^{1+\gamma}  \; ,
 \end{equation}  
where $F_n$ is the diffusive particle flux at $L = L_0 =  12$ (with a value corresponding to $330$ kg s$^{-1}$ 
used in \citep{Saur04}), 
$F_E$ is the radial energy flux at $L_0$ and is treated as a fitting parameter.   
The heating rate function $q$ calculated in \citep{Saur04} is based on
a weak turbulence theory \citep{Ng96} using {\it Galileo} data.  Based on this $q$ function,  \citet{Saur04} 
found that the observed increasing plasma temperature profile could be explained
through inward diffusive transport of thermal energy of $F_E = -2.0$ TW 
generated in the middle
magnetosphere by  turbulent heating (e.g., 20-30 ${\rm R_J}$) 
with an outward energy flux at $L = 35$.   

In reconsidering the turbulent heating process, we have also re-analyzed the heating rate
function $q$ as presented in Sections~\ref{turb-heat-model} and \ref{data-analy}.
Motivated by the significant difference between our calculation and what was calculated by 
\citet{Saur04}, 
we look into Eq.~(\ref{equ:temp-integrated}) more carefully. 
We see that the qualitative trend of an increasing temperature resulted from Eq.~(\ref{equ:temp-integrated}) is in fact not very sensitive to the magnitude of $q$. 
In fact, the increasing trend is mainly determined by the dependence on $L$ in the first
and third terms.
In the large $L$ limit, both terms are proportional to $L^{\beta - 3 -\gamma}$, or $L^{1.6}$ 
based on parameters used in the paper, and thus are increasing roughly at the same rate
as in observed data.
The $L^{\beta - 3 -\gamma}$ dependence is originated from the adiabatic condition
as can be inferred also from Eq.~(\ref{equ:e-diffusion}). 
In \citep{Saur04}, the special case of $q = 0$ is considered. 
For this case, $F_E$ has to be chosen as $T_0 F_n$ so that both the first and third terms
are identically zero. This would result in a decreasing temperature dependence of $L^{-\gamma}$,
or $L^{-2}$, 
due to the second term. 
However, for any non-zero $q$ such that $F_E \ne T_0 F_n$ exactly, 
the temperature will be increasing due to the first and the third terms since the second term 
becomes small for larger $L$. 
Eq.~(\ref{equ:temp-integrated}) is therefore not sensitive enough to constrain the magnitude of $q$ 
based on an observed increasing profile.
Nevertheless, turbulence can still
play an important role in heating Jupiter's magnetodisc.  
As discussed above,  it is actually reasonable to approach the heating
problem with an advective outflow model, similar to the application in the solar
wind heating problem. 
With this approach, we show in the following sections that the
observed heating is consistent with dissipation of turbulent magnetic
field fluctuations. 

\section{Turbulent plasma heating}

\subsection{Temperature model}
			
A turbulent heating model based on advective transport can be derived
from the heating equation of an ideal gas,
\[
\frac{3}{2}n^{5/3}\frac{d}{dt}\left(pn^{-5/3}\right)=q \, , 
\]	
where $n$ is the number density, $p=nk_BT$ is
the pressure, and $q$ is the heating rate per
volume. The time derivative is taken as the
convective derivative $d/dt =
\partial/\partial t+{\bf v}\cdot {\bf \nabla} =
{\bf v}\cdot {\bf \nabla}$ for steady state, where ${\bf v}$ is
the advective velocity, with a magnitude $v$, so that 
\[
v\frac{d}{dr}\left(k_BTn^{-2/3}\right)=\frac{2}{3}n^{-5/3}q \, .
\]
Using $r=LR_J$ and assuming $n \propto L^{-\beta}$ as in \citep{Saur04}, this becomes
\[
\frac{dT}{dL}=-\frac{2\beta T}{3L}+\frac{2R_J}{3k_Bnv}q \, .
\]
The advective velocity is assumed to carry out in steady state the mass loading rate $\dot{M}$ (taken to be 330 kg/s). 
Note that due to magnetic flux conservation there must be return flow from the outer 
region back to the inner region. 
However, since in MHD average velocity is weighted by mass density, 
the outflow carrying mass dominates in the average over the return flow after mass 
is dumped in the outer magnetosphere.
Moreover, the outflow with higher density also contributes more in the averaging of
ion temperature.
Therefore we assume in this paper that the increase in ion temperature is mainly 
due to the heating following the outflow.
If the plasma is assumed to be confined to a disk thickness $\sim HR_J$, where 
$H$ is the scale height, then $\dot{M} \sim 2\pi LHR_J^2m_inv$ so that $nv \propto L^{-1}$ and the temperature equation can then be written as
\begin{equation}
  \frac{dT}{dL}=-\frac{2\beta T}{3L}+c_1 L^\alpha q \, ,
  \label{dT_dL_equation}
\end{equation}	
where $\alpha$ and $c_1$ are parameters determined by confinement geometries. In the case where plasma is confined to a disk $\alpha = 1$ and $c_1$ can be written as
\[
c_1 = \frac{2}{3}\frac{R_J^32\pi m_iH}{\dot{M}k_B}  \, ,   
\]
so that it is about $2 \times 10^{19}~{\rm Km^3}/{\rm W}$. 
Eq.~(\ref{dT_dL_equation}) can then be solved for temperature,
\begin{equation}
  T=T_0\left(\frac{L}{L_0}\right)^{-2\beta/3}+c_1\left(\frac{L}{L_0}\right)^{-2\beta/3}\int_{L_0}^{L}q(L')\frac{L'^{\alpha+2\beta/3}}{L_0^{2\beta/3}}dL' \, .
  \label{T_equation}
\end{equation}	
The first term in this temperature equation satisfies the adiabatic condition in the absence of heating and depicts 
cooling of the expanding plasma. The second term describes the effect of turbulent heating. 

\subsection{Plasma turbulent heating model}
\label{turb-heat-model}
In a turbulent theory \citep{Kolmogorov41, Iroshnikov63, Kraichnan65, Ng96}, 
the energy cascade rate from one scale to another is given by
\[
 \epsilon \sim \frac{E_{k}}{\tau} \, ,
\]
where the $E_{k}$ is the energy and $\tau$ is the transfer time scale at a spatial scale
(wavenumber) $k$. 
Assuming equipartition between kinetic and magnetic energy in MHD turbulence, we have
\[
E_{k} \sim 2\frac{\delta B_\perp^2}{2\mu_0\rho} \, ,
\]
where $\delta B_\perp$ is the magnitude of magnetic field fluctuations at scale $k$ 
perpendicular to the large scale magnetic field.
The dispersion relation of wave packets is
$\omega \sim k_\parallel v_A \sim v_A/\lambda_\parallel$,
where $v_A$ is the Alfv\'{e}n speed
and $\lambda_\parallel$ is the spatial scale parallel to the large scale
magnetic field at scale $k$. The displacement $\delta r$ of the field per each interaction can be estimated from
the Walen relation (also valid for nonlinear Alfv\'{e}n waves)
\[
\frac{\delta u_\perp}{v_A} = \frac{\delta r \delta t}{\delta t\lambda_\parallel}=\frac{\delta B_\perp}{B_0} \, ,
\]
where $\delta t$ is the interaction time and 
$B_0$ is the magnitude of the large scale magnetic field, so that
\[
\delta r = \frac{\delta B_\perp}{B_0} \lambda_\parallel \, .
\] 
The fractional change as compared with the perpendicular scale $\lambda_\perp$ 
of the wave packet is
\[
\chi \sim \frac{\delta r}{\lambda_\perp} \sim \frac{\delta B_\perp}{B_0} \frac{\lambda_\parallel}{\lambda_\perp} \sim \frac{\delta B_\perp}{B_0} \frac{k_\perp}{k_\parallel} \, . 
\]
When $\chi \ll 1$, the wave packets are only slightly altered by the distortion of the magnetic field line
in one interaction time and thus for weak turbulence many random interactions are required to induce 
a fractional change or order unity \citep[]{Ng96}. 
The turbulent cascade time scale is therefore
\[
\tau \sim \chi^{-2} \frac{1}{k_\parallel v_A} \sim \sqrt{\mu_0 \rho}\frac{B_0}{\delta B_\perp^2}\frac{k_\parallel}{k_\perp^2} \, .
\]
Then the turbulent heating rate of plasma can be found from the energy cascade rate
\begin{equation}
  q_{\rm MHD-weak} \sim \epsilon \rho \sim \frac{1}{\sqrt{\mu_0^3 \rho}}\frac{\delta B_\perp^4}{B_0}\frac{k_\perp^2}{k_\parallel} \, .
  \label{MHD_eq}
\end{equation}
In the strong turbulence limit $\chi \rightarrow 1$ and the Walen relationship becomes $k_\parallel v_A \rightarrow k_\perp \delta u_\perp$
so that the turbulent cascade time becomes
\[
\tau \sim \frac{1}{k_\perp \delta u_\perp} \sim \frac{\sqrt{\mu_0 \rho}}{ k_\perp \delta B_\perp} \, ,
\]
and the heating rate density is given by
\begin{equation}
  q_{\rm MHD-strong} \sim \frac{\delta B_\perp^3 k_\perp }{\sqrt{\mu_0^3 \rho}} \, .
  \label{MHD_strong_eq}
\end{equation}	

Spectral indices corresponding to cascade rates based on Eq.~(\ref{MHD_eq}) or 
Eq.~(\ref{MHD_strong_eq}) are $-2$ and $-5/3$ respectively. 
As discussed in more detail later, observed spectral indices are mostly 
between these two values, as shown in Fig.~\ref{slopes_hist_fig}. 
In this paper, we will present calculations based on Eq.~(\ref{MHD_eq}) only.
We have also used Eq.~(\ref{MHD_strong_eq}) in calculations which result in
stronger heating but with qualitatively similar trend. 
Therefore, we will not repeat
the discussion for that case here.

\section{Data analysis}
\label{data-analy}

Turbulent heating processes are estimated from Galileo magnetometer
(MAG) \citep{Kivelson-etal-92} 
observations in the
jovian magnetosphere. The magnetic field was analyzed in spherical Jupiter
centered solar magnetic coordinates. The radial coordinate
$\widehat{\boldsymbol{e}}_r$ is in the direction away from the
planet. Azimuthal coordinate is in the direction of corotation such
that $\widehat{\boldsymbol{e}}_\phi$ is perpendicular to the plane
defined by $\widehat{\boldsymbol{e}}_\Omega$ and $
\widehat{\boldsymbol{e}}_r $ where $\widehat{\boldsymbol{e}}_\Omega$
is in the direction of the magnetic dipole axis, and
$\widehat{\boldsymbol{e}}_\theta = \widehat{\boldsymbol{e}}_\phi
\times  \widehat{\boldsymbol{e}}_r$ completes the right hand coordinate
system.  

For this analysis two-hour windows with at least 600 measurements
(i.e. average $\Delta t \leq 12 \,{\rm s}$) were selected. Linear interpolation
was then used to break the time series into regular sampling
intervals. The window mean magnetic field $\boldsymbol{B}_0(t)$ was chosen
as a 2000 s moving average, so that the size of the analyzed window
is a little over 5000 s (with small variations depending on the
original sampling rate). 
		 		
Perturbation of the magnetic field is then calculated as $\delta
\boldsymbol{B}(t) = \boldsymbol{B}(t) - \boldsymbol{B}_0(t)$, and
perpendicular fluctuations of the field as $\delta
\boldsymbol{B}(t)_\perp  =  \delta \boldsymbol{B}(t) - \delta
\boldsymbol{B}(t)_\parallel$, where $\delta
\boldsymbol{B}(t)_\parallel$ is the component of the perturbation in
the direction of $\boldsymbol{B}_0(t)$. The power spectrum of the vector
components of $\delta \boldsymbol{B}(t)_\perp$ was then estimated as
\citep[]{Tao15}	
\[
P(f) = \frac{2}{N\Delta
  t}\sum_{i=1}^{N}\Delta t \left|W_i(t_i,f)\right|^2 \, ,
\] 
with the
wavelet period of $(1.03f)^{-1}$ \citep[]{Farge92, Torrence98}. The
total power spectrum of the perpendicular fluctuation was 
calculated as a square root of the sum of squares of the component power spectra.  
		
\begin{figure}
  \noindent\includegraphics[width=30pc]{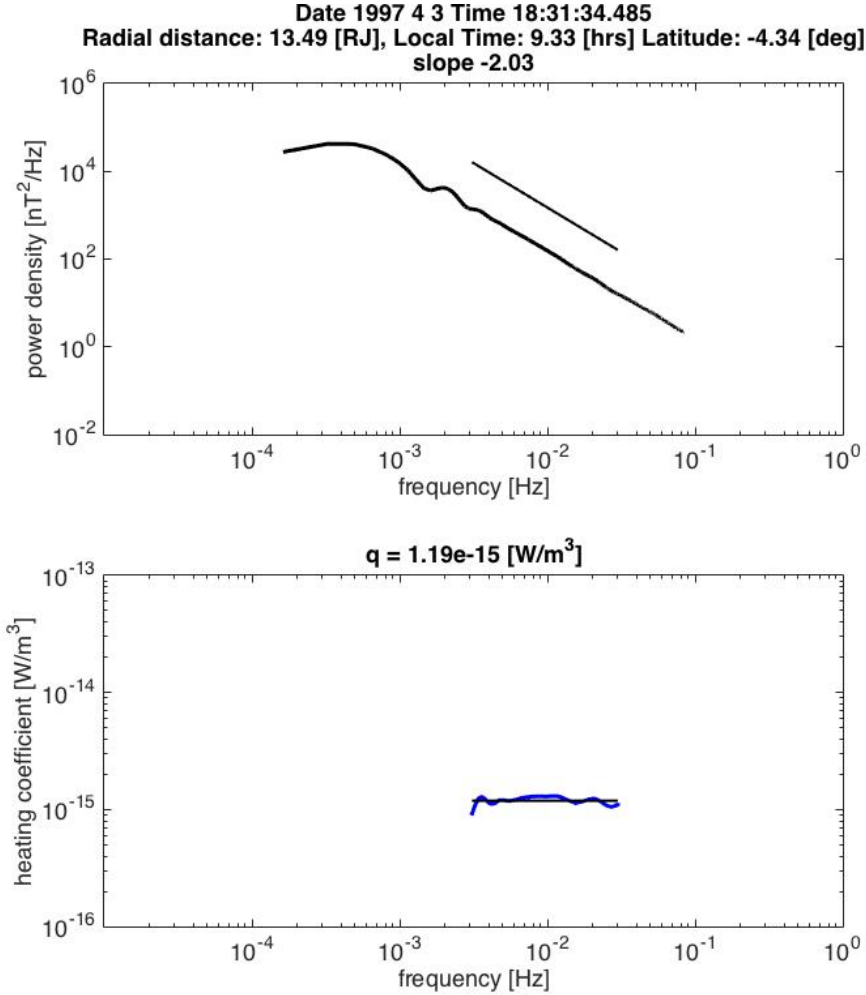}
  \caption{An example of the calculation of the heating rate density (bottom panel) from a power
    spectrum (top panel). A weak turbulence model is used to calculate the heating
    rate density in the frequency subrange [$3 \times 10^{-3}$, $3
    \times 10^{-2}$] Hz. The straight line in the top panel is to indicate a slope of $-2$ over this range
    as expected 
    from the weak turbulence model. The straight line in the bottom panel indicates an average
    value.} 
  \label{power_spec_fig}	
\end{figure}

The heating rate density was calculated from the slope of the power
spectrum of $\delta \boldsymbol{B}(t)_\perp$ in the range of
frequencies between $3 \times 10^{-3}$ Hz and $3\times 10^{-2}$ Hz or half of the ion gyro
frequency $f_i = ZeB_0/m_i$ (whichever is smaller). Using
conservation of energy $\delta b^2 \sim P(f)f$ \citep[]{Leamon99}
and Eq.~(\ref{MHD_eq})
the heating rate density can be calculated as
\begin{equation}
  q = \frac{1}{\sqrt{\mu_0^3 \rho}}\frac{\delta b^4}{B_0}\frac{k_\perp^2}{k_\parallel} \, ,
  \label{MHD_weak_calc_eq}
\end{equation}
where $k_\parallel = (HR_J)^{-1}$,  
$ k_\perp = 2 \pi f/v_{\rm rel} \sin (\theta_{vB})$ \citep[]{vonPapen14}
with $v_{\rm rel}$ being the magnitude of the flow velocity in the spacecraft frame and
$\theta_{vB}$ being the angle between the flow velocity and $\boldsymbol{B}_0$, 
and $B_0$ is the magnitude of the time average of
magnetic field over the analyzed window. Empirical radial profiles of
scale height $H$ and plasma density
in the magnetodisc is given by \citet{Bagenal11}. Plasma
in the jovian magnetosphere is composed mainly of oxygen and sulfur
ions, expelled by volcanic activity on the moon Io.  The average ion mass
used in this analysis is taken to be 22 AMU.  The heating rate
density was calculated and averaged over the analyzed frequency range.
An example of this calculation is shown in Figure \ref{power_spec_fig}. 
For this analysis we use samples where
the heating rate density varies less than a factor of 15 over the whole range, 
and we limit the relative magnetic field fluctuations to
$\delta B_\perp  < 0.2 B_0$ with the intent of removing larger
fluctuations that could potentially be associated with spatial structures.

\begin{figure}
  \noindent\includegraphics[width=30pc]{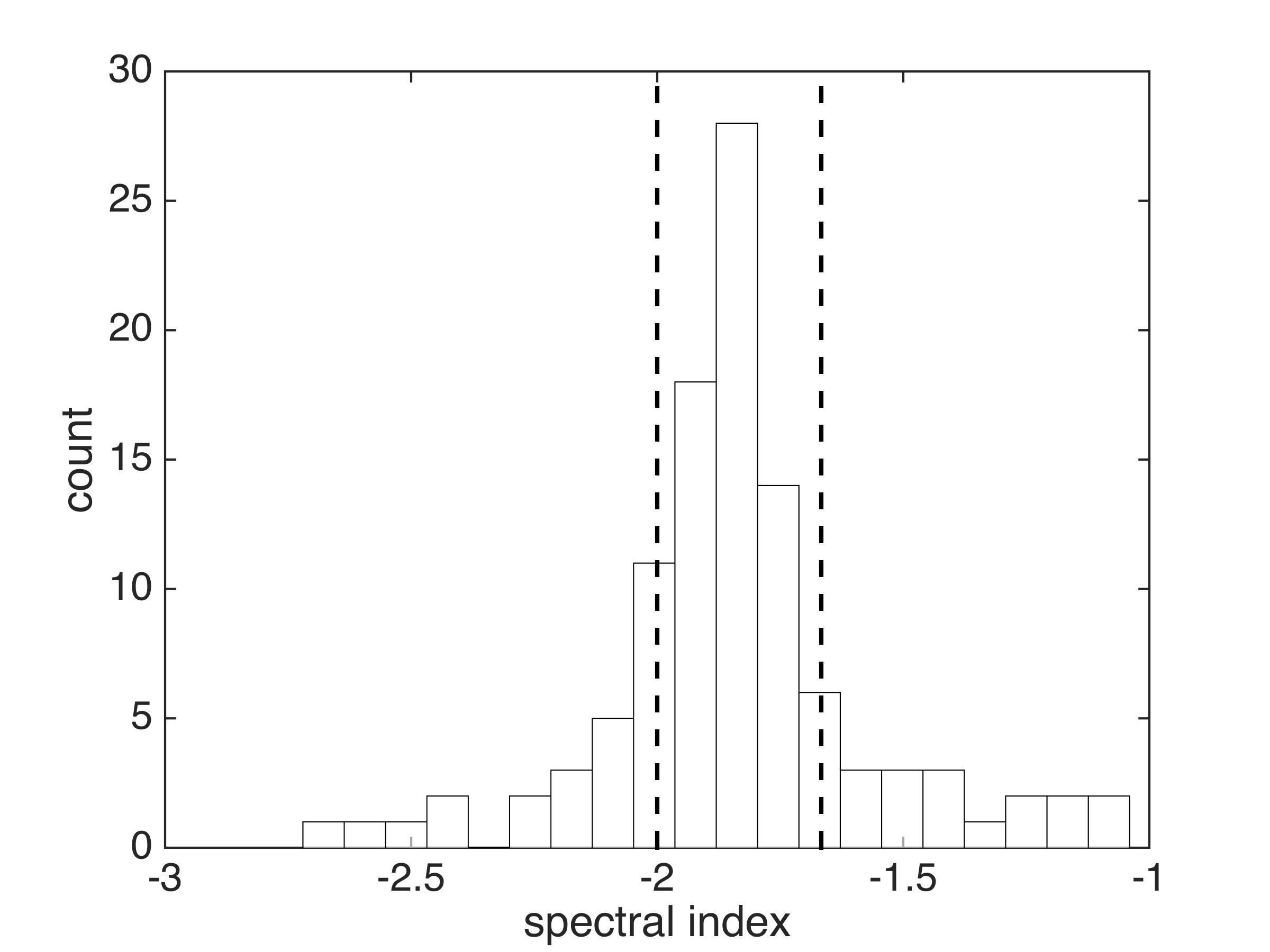}
  \caption{Histogram of slopes of power spectra. The majority of values
    lie between $-5/3$ and $-2$ (represented by dashed lines).} 
  \label{slopes_hist_fig}	
\end{figure}

\begin{figure}
  \noindent\includegraphics[width=30pc]{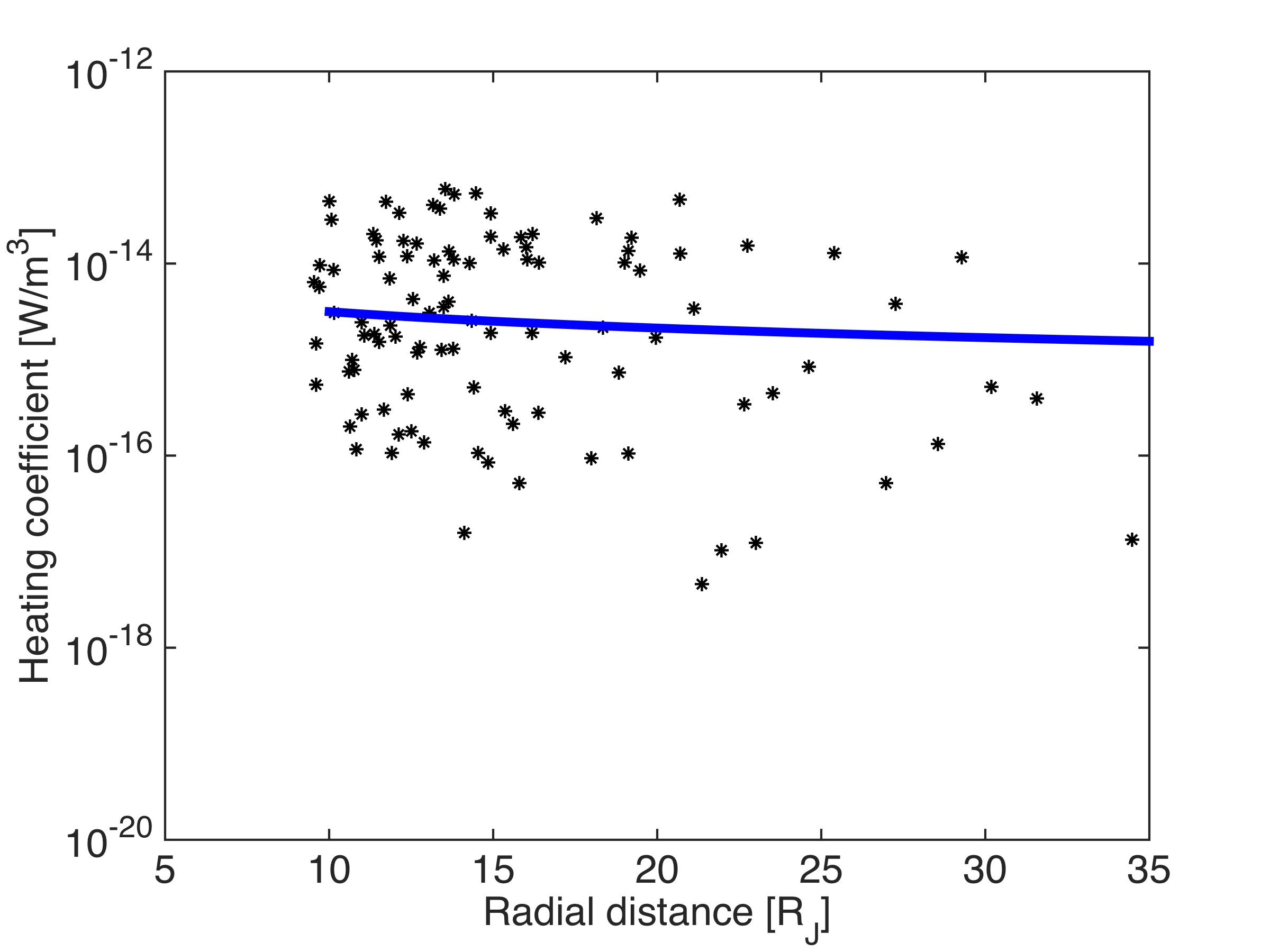}
  \caption{Heating rate density  as a function of the radial
    distance in Jupiter radii. The blue line depicts a power law,
    $q~\approx~(1.2\times 10^{-14})L^{-0.57}$ W$/$m$^3$,
     fit to
    the data averaged in $1~{\rm R_J}$ bins.} 
  \label{q_ave_fig}	
\end{figure}

Employing the selection criteria described above, 108 samples were
chosen. A histogram of the slopes of power spectra in the analyzed
frequency range of the selected samples is presented in Figure
\ref{slopes_hist_fig}, with the peak of values between $-1.5$ and
$-2$. Figure \ref{q_ave_fig} shows the turbulent heating rate density between
10 and 35 ${\rm R_J}$.  A power law fit to
heating rate density averaged over 2${\rm R_J}$ bins using a geometric
mean has the form $q \approx (1.2 \times 10^{-14})L^{-0.57}$ W$/$m$^3$ (blue
curve). 

Note that our heating rate density is significantly different from what is shown in 
\citet{Saur04}, especially for $L < 20$.
This is due to the fact that \citet{Saur04} imposes an absolute limit of 
the magnetic fluctuations ($\delta B_\perp  < 0.7 \, {\rm nT}$)
for all range of $L$ \citep{saur-etal-2002}, as compared with our relative 
limit of $\delta B_\perp  < 0.2 B_0$.
This means that for smaller $L$, \citet{Saur04} excludes more data with stronger fluctuations 
than our calculation
since $B_0$ is larger for smaller $L$, and thus results in smaller heating rate density.
This difference is important since the temperature calculation presented below 
depends critically on our new calculation. 
If a much smaller heating rate density is used in the temperature calculation,
the temperature would not be increasing fast enough to be consistent with observations.

\begin{figure}
  \noindent\includegraphics[width=30pc]{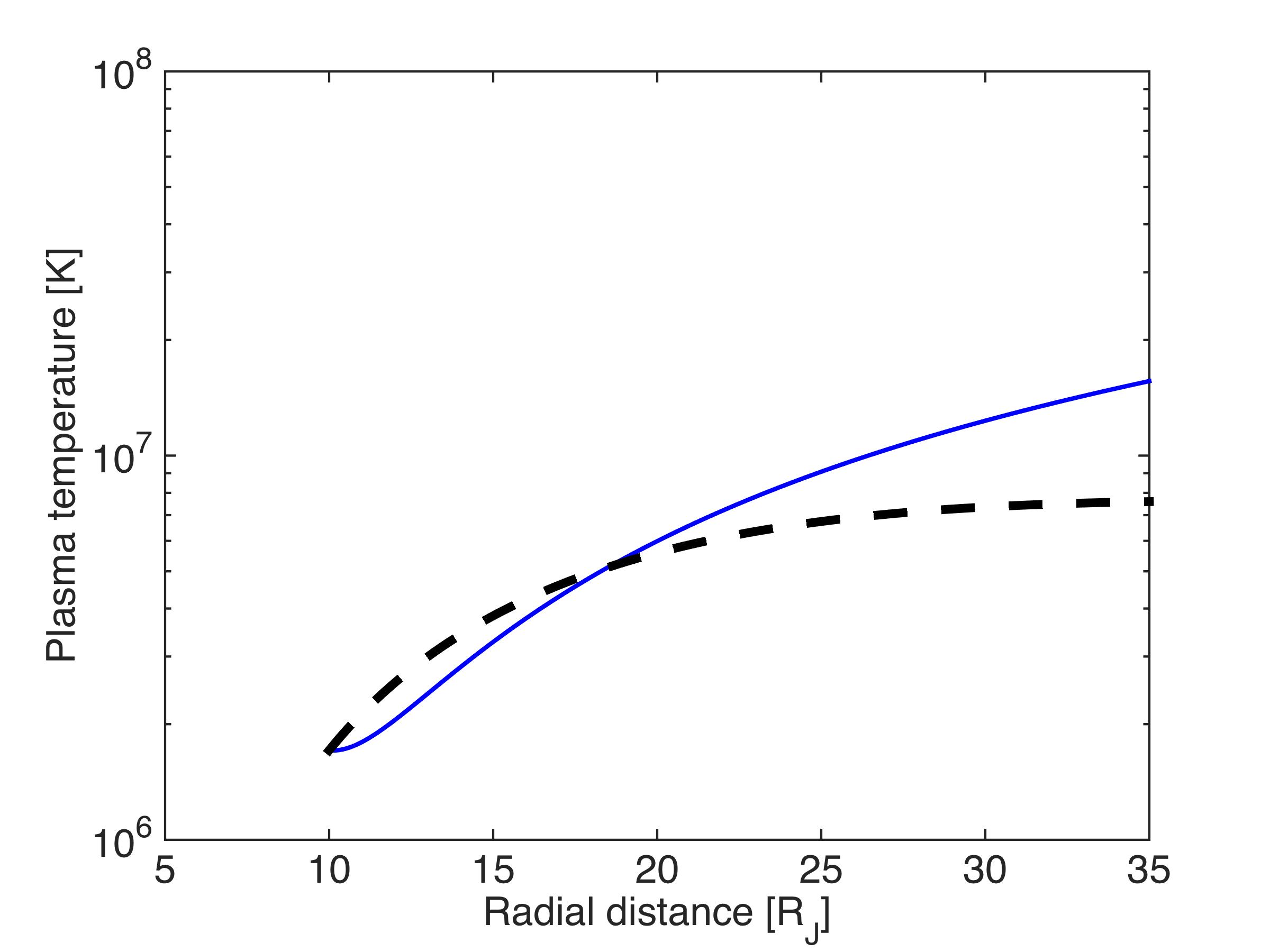}
  \caption{Temperature profile (blue line) calculated from Eq.~\ref{T_equation}, 
  using the calculated turbulent heating rate density shown in
    Figure \ref{q_ave_fig}. Black dashed line
    represents a temperature profile from \citet{Bagenal11}. } 
  \label{temperature_fig}	
\end{figure}

Using the heating rate density fit and Eq.~(\ref{T_equation}),  
temperature can then be calculated as a function of
the radial distance as shown in  Figure \ref{temperature_fig} (blue curve). Here
the initial temperature $T_0$ was taken to be $1.7 \times 10^6$ K ($\sim$150 eV) from
the empirical temperature profile of
\citet{Bagenal11} (black dashed curve).  Within a factor of two, the
profiles are in general agreement.

\section{Discussion}

Understanding the physical mechanisms that lead to plasma heating in
the giant magnetospheres is a decades-long conundrum.  Magnetodisc
equilibrium models (e.g., \citep{Caudal86}) have demonstrated the role
of plasma pressure and even pressure anisotropy in radial stress
balance \citep[]{Paranicas91}.  An obvious energy source is planetary
rotation and the centrifugal potential \citep[]{Vogt14}, though the
solar wind can also contribute \citep[]{Bagenal11}.  In this paper, we
have explored the role of turbulent magnetic fields as a plasma
heating mechanism, following \citet{Saur04}. 
However, we have adopted an
alternative transport model based on 
advective outflow of magnetodisc plasma  beyond 10
${\rm R_J}$, showing that the heating rate density due to turbulent
dissipation is sufficient to account for Jupiter's observed radial ion
temperature profile.  

Recent observations by the Juno's Jupiter
Energetic-particle Detector Instrument (JEDI) \citep[]{Mauk17a, Mauk17b} suggest
that auroral electrons at Jupiter are dominated by power-law
distributions.  These broad-band energy distributions are suggestive
of energization by dispersive scale Alfv\'{e}n waves
\citep[e.g.][]{GRL:GRL15179, JGRA:JGRA16623, Wing13}
often attributed to Alfv\'{e}nic
aurora at Earth
where Alfv\'{e}nic energy may reach dispersive scales via a turbulent cascade
\citep{PhysRevLett.100.175003}.  
In this case, we infer a connection
between high latitude acceleration by, e.g., inertial Alfv\'{e}n waves
and the observation of kinetic Alfv\'{e}n waves (i.e., strong magnetic field
fluctuations) in the equatorial
magnetodisc.  
Alfv\'{e}n waves on these dispersive scale lengths \citep[]{JGRA:JGRA12870} are capable of
converting significant
Poynting flux to electron kinetic energy (precipitation)
\citep[]{GRL:GRL15179, GRL:GRL17124, doi:10.1063/1.2744226}.   
Likewise, following \citet{Saur04}, we infer
that kinetic Alfv\'{e}n waves in the magnetodisc can serve as the catalyst in ion heating
\citep{jrj:2001b} to complete the turbulent cascade.  

An interesting difference between
Jupiter and Earth is the role of multiple resonant cavities due to
density variations along the magnetic field line \citep[]{Delamere16}.  Wave transmission
to high latitude is a function of parallel wavelength, and
\citep[]{Wright89, Delamere03_Io, Hess10} showed that significant reflection
can occur for perturbations associated with the Io-Jupiter
interaction.  The non-linear interaction between counter propagating
waves leads to a turbulent cascade.  We suggest here that resonant
cavities could inhibit steady-state magnetosphere-ionosphere coupling
currents and, in fact, promote turbulence.  

An additional consideration for the middle and outer magnetosphere are
the long Alfv\'{e}n travel times.  \citet{Bagenal07} showed that travel
times can approach the order of one hour, which is a non-negligible
fraction of the planetary rotation period ($\sim$10 hr).   If
steady M-I coupling currents are prohibited due to the inability of the
system to promptly respond to fluctuations (driven by, for example,
local time variations), then M-I decoupling is mandatory.  Parallel
electric fields facilitate decoupling by breaking the frozen-in
condition, and we note that parallel electric fields are an inherent
property of kinetic/inertial Alfv\'{e}n waves.   

Jupiter is certainly not unique.  Turbulent heating can account for
magnetodisc heating at Saturn too. \citet{Kaminker17}, 
\citet{vonPapen14}, and \citet{vonPapen16} showed that
magnetic field fluctuations measured by the Cassini magnetometer (MAG)
instrument are consistent with the requisite turbulent heating rate
density \citep[]{Bagenal11}.   Using the 1s-average MAG data, \citet{Kaminker17} compared the heating rate density in both the inertial subrange
(MHD scale) and the dissipation scale (kinetic scale) and found that
the kinetic scale heating was typically larger.  An energy-conserving
cascade would predict equal values in both subranges, so the question
remains whether energy could be injected at the kinetic scale via, for
example, magnetodisc reconnection \citep[]{Delamere15}.   A similar analysis that resolves
the kinetic time scale using Juno data should be conducted in future studies.

\section{Summary}

We have analyzed Galileo magnetometer data to investigate plasma heating by turbulent magnetic field fluctuations
using an advective transport model.
We summarize our findings as follows:
\begin{itemize}
\item Our advective outflow model for investigating turbulent heating (e.g., appropriate for the solar
  wind) is different from the previous
  studies.  \citet{Saur04} used a diffusive transport model.  We argue
  that an advective outflow approach is reasonable beyond 10 ${\rm
    R_J}$ where transport becomes rapid and dominated by large-scale
  motion.

\item We re-calculate the heating rate density and obtain much higher values 
for $L < 20$ than what is used in \citet{Saur04}. This increase is critical in the calculation
of temperature that is consistent with observation data.

\item Using an inner boundary at 10 ${\rm R_J}$ and specifying an
  inner boundary temperature (150 eV), we find that Jupiter's radial ion
  temperature profile is consistent with heating by turbulent magnetic
  field fluctuations. 

\item Turbulence appears to be ubiquitous in the giant magnetospheres
  and could have significant implications for
  magnetosphere-ionosphere coupling and auroral processes. 

\end{itemize}

\acknowledgments
The authors acknowledge support from NASA grants NNX14AM27G, NNX15AU61G,
and NNH15AZ95I.  The Galileo magnetometer data used in this analysis was obtained from the
Planetary Data System (http://pds.nasa.gov/).

\listofchanges

\end{document}